# Accelerated discovery and design of Fe-Co-Zr magnets with tunable magnetic anisotropy through machine learning and parallel computing

Weiyi Xia[1,2], Maxim Moraru[3], Ying Wai Li[3], Timothy Liao[4,5], James R. Chelikowsky[4,5,6] and Cai-Zhuang Wang[1,2]

[1]*Ames National Laboratory, U.S. Department of Energy, Iowa State University, Ames, Iowa 50011, USA*
[2]*Department of Physics and Astronomy, Iowa State University, Ames, Iowa 50011, USA*
[3]*Computer, Computational, and Statistical Sciences Division, Los Alamos National Laboratory, Los Alamos, NM 87545, USA*
[4]*Center for Computational Materials, Oden Institute for Computational Engineering and Sciences, The University of Texas at Austin, Austin, Texas 78712, USA*
[5]*Department of Physics, The University of Texas at Austin, Austin, Texas 78712, USA*
[6]*McKetta Department of Chemical Engineering, The University of Texas at Austin, Austin, Texas 78712, USA*

**Abstract**
Rare earth (RE)-free permanent magnets, as alternative substitutes for RE-containing magnets for sustainable energy technologies and modern electronics, have attracted considerable interest. We performed a comprehensive search for new hard magnetic materials in the ternary Fe-Co-Zr space by leveraging a scalable, machine learning-assisted materials discovery framework running on GPU-enabled exascale computing resources. This framework integrates crystal graph convolutional neural network (CGCNN) machine learning (ML) method with first-principles calculations to efficiently navigate the vast composition-structure space. The efficiency and accuracy of the ML approach enable us to reveal 9 new thermodynamically stable ternary Fe-Co-Zr compounds and 81 promising low-energy metastable phases with their formation energies within 0.1 eV/atom above the convex hull. The predicted compounds span a wide range of crystal symmetries and magnetic behaviors, providing a rich platform for tuning functional properties. Based on the analysis of site-specific magnetic properties, we show that the $Fe_6Co_{17}Zr_6$ compound obtained from our ML discovery can be further optimized by chemical doping. Chemical substitutions lead to a ternary $Fe_5Co_{18}Zr_6$ phase with a strong anisotropy of $K_1$ = 1.1 MJ/m$^3$, and a stable quaternary magnetic $Fe_5Co_{16}Zr_6Mn_4$ compound.

## I. INTRODUCTION

High-performance permanent magnets are critical to the advancement of modern technologies including renewable energy, transportation, and information storage. For decades, rare-earth-based magnets, such as $Nd_2Fe_{14}B$ [1] and $SmCo_5$ [2], dominate many of these applications due to their exceptional magnetic properties. However, concerns about the supply, cost, and environmental impact of rare-earth elements have intensified the search for rare-earth-free alternatives that can deliver comparable performance [3], particularly in terms of high magnetic polarization and strong magnetocrystalline anisotropy—key attributes for efficient and robust permanent magnets [4-6].

Transition metals, especially Fe and Co, are attractive constituents for magnets owing to their abundance and large atomic magnetic moments. Nevertheless, binary Fe-Co alloys typically crystallize in cubic structures, resulting in low magnetic anisotropy that limits their effectiveness as permanent magnets. Binary or ternary compounds formed by combining Fe, Co, or FeCo with heavier transition metals such as Zr, Hf, and Nb have shown to exhibit significant anisotropies, although often with a trade-off in magnetization [7-13]. In particular, Zr is a well-established stabilizer for complex intermetallic structures in transition-metal-rich alloys and has been shown to induce noncubic symmetries, which are essential for enhancing magnetocrystalline anisotropy [8-13]. Recent studies have shown that Zr substitution in Fe, Co, and FeCo-based compounds can improve phase stability, refine microstructures, and significantly enhance magnetic anisotropy and coercivity—effects attributed to both electronic and structural modifications introduced by Zr [8-13]. These studies have highlighted the role of Zr in stabilizing novel magnetic phases, motivating a systematic exploration of the Fe-Co-Zr system for rare-earth-free permanent magnet applications.

Despite these advances, the vast compositional and structural space of multinary systems poses a formidable challenge for conventional experimental and computational approaches. The number of possible combinations and crystal structures increases exponentially with each added element, making exhaustive trial-and-error synthesis or brute-force computational searches impractical. Recent progress in machine learning (ML) and high-throughput first-principles calculations has revolutionized the materials discovery process. ML models, trained on large datasets of first-principles calculations, can rapidly predict formation energies and magnetic properties across hundreds of thousands of candidate structures, efficiently narrowing the search space to the most promising compositions [14-29]. These candidates are then validated and refined through targeted first-principles calculations, enabling the identification of stable and metastable compounds with tailored magnetic functionalities.

In this context, the Fe-Co-Zr system represents a fertile ground for the discovery of rare-earth-free magnetic materials with tunable anisotropy and high magnetization. Leveraging state-of-the-art ML-guided frameworks, this study systematically explores the Fe-Co-Zr compositional space, leading to the prediction and characterization of new stable and metastable compounds with unique magnetic properties. We report the discovery of 9 stable and 81 metastable new Fe-Co-Zr ternary compounds. Furthermore, we demonstrate a novel approach for optimizing magnetic properties via targeted local atomic substitutions, significantly enhancing the magnetic anisotropy of a promising candidate. Our approach not only accelerates the pace of discovery but also provides fundamental insights into the structure-property relationships underpinning high-performance, rare-earth-free magnets.

## II. COMPUTATIONAL FRAMEWORK AND METHODS

**exa-AMD, an ML-Guided High-Throughput Screening framework** - The core of our high-throughput screening is the exa-AMD, an ML-guided high-throughput screening framework[30]. It is a Python-based application designed to accelerate materials discovery by integrating AI/ML tools and quantum mechanical calculations into a scalable workflow. The execution is managed by Parsl [31], a parallel programming library that allows tasks to be flexibly executed on resources ranging from laptops to supercomputers. The advantages of this exa-AMD framework lie in its efficiency and scalability— enabling massive screening campaigns—and its elasticity, as computing resources like GPUs can be dynamically allocated or released to match task requirements. The workflow is modular and configurable. It allows users to resume computations from intermediate steps and configure execution settings to balance performance and accuracy. A schematic of the ML-guided exa-AMD workflow for searching ternary Zr-Fe-Co compound is shown in **Fig. 1**. The workflow proceeds in five main stages as described below.

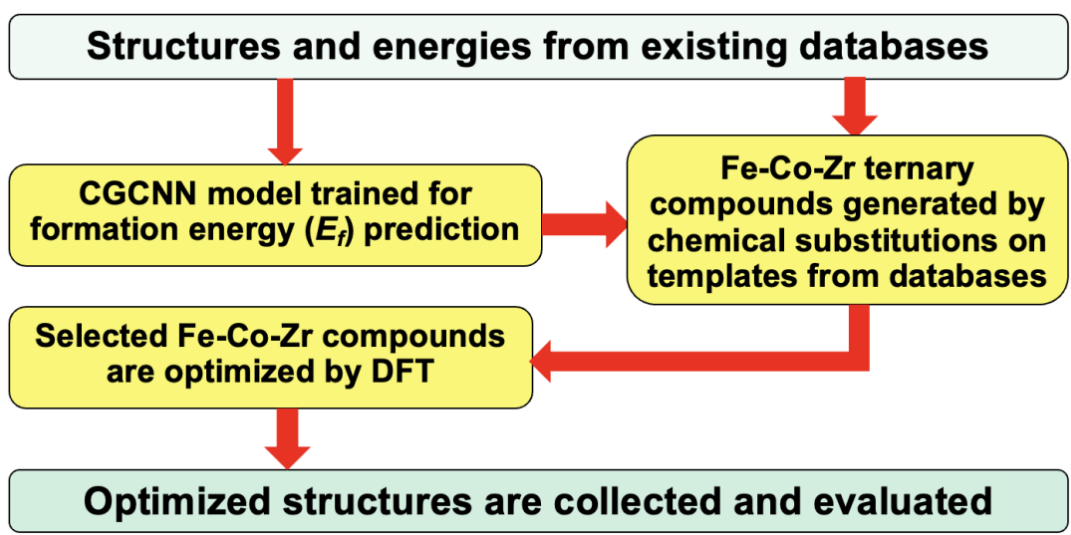

**Fig. 1.** A schematic of the ML-guided exa-AMD workflow for the discovery of ternary Zr-Fe-Co compounds. The key computational stages, highlighted in yellow, are all highly parallelized to run efficiently on either CPU or GPU resources.

**(i) Structure Generation-** A vast pool of hypothetical Fe-Co-Zr structures is generated by substituting Fe, Co, and Zr atoms onto the lattice sites of known ternary crystal structure prototypes obtained from materials databases. By extracting 28472 crystalline ternary structures from the Materials Project (MP) database [32], as well as 1013 unique magnetic materials from our Novomag database [33], we substitute the three elements in these known ternary compounds with Fe, Co and Zr respectively. The template structures from MP cover most of the currently known structure motifs. We apply no discrimination in the structure motifs when extracting these template structures from the MP database to ensure unbiased ML screening in identifying low-energy structures for the systems of interest. For each ternary template structure from MP and Novomag, we generate 30 structures by shuffling the order of the three elements (6 ways) and by uniformly expanding or contracting the volume of the structure by 5 scaling factors (0.92, 0.96, 1.0, 1.04, and 1.08). The reason for varying the volume of the template structure is that the sizes of the elements in the template structure would be very different from those in our Fe-Co-Zr systems, thus the corresponding equilibrium volumes would be very different. This practice helps capture better candidates in the subsequent ML screening. In this way, a structure pool of 884550 hypothetical ternary compounds covering a wide range of compositions is generated for each Fe-Co-Zr system. No structure relaxations are performed for these hypothetical structures before the ML screening.

**(ii) Formation Energy Prediction:** The generated structures are rapidly screened using a Crystal Graph Convolutional Neural Network (CGCNN) model, which predicts the formation energy of each candidate. CGCNN represents a crystal structure as a graph, where atoms are nodes and interatomic bonds are edges, allowing it to efficiently learn composition-structure-property relationships. It has been demonstrated that CGCNN ML models can be trained to efficiently evaluate the formation energies ($E_f$, see the definition below) of compounds [17]. The universal CGCNN model is trained on a broad set of compounds from the MP database. CGCNN is selected as the ML model in our study due to its demonstrated ability to predict formation energies of inorganic compounds. While newer models such as ALIGNN [34] or E3NN [35] offer advantages in certain applications, the balance of simplicity, interpretability, and efficiency in CGCNN makes it well-suited for this study. With GPU acceleration, the formation energy prediction of 1 million structures takes less than 15 minutes using a single GPU node (4 A100 GPUs).

**(iii) Structure Selection and Filtering:** By applying the CGCNN ML model to screen the formation energy of the 884550 hypothetical ternary compounds, structures with low predicted formation energies are selected as promising candidates. In this work, we apply a criterion of $E_f$ < -0.1 eV/atom for the selection. The formation energy distribution (histogram) from the CGCNN predictions is shown in **Fig. 2**. Based on the $E_f$ histogram shown in **Fig. 2**, a filtering stage is then applied to remove duplicate or structurally similar candidates. We down selected 3102 structures through this process, which are ~0.35% of the structures screened by CGCNN, for further evaluation by first-principles calculations.

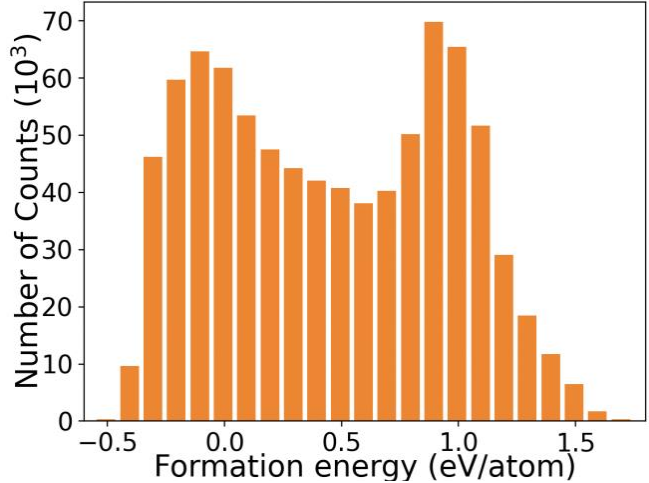

**Fig. 2.** Distribution of formation energies ($E_f$) of the substitutional ternary Fe-Co-Zr ternary compounds predicted from the CGCNN energy model. The total number of structures are 884550. Structures with $E_f$ less than -0.1 eV/atom from this prediction are selected for further analysis.

**(iv) First-Principles Calculations:** The set of 3102 promising structures selected by the CGCNN screening is passed to the next stage for validation using first-principles calculations based on density functional theory (DFT). All first-principles calculations were performed using the Vienna Ab initio Simulation Package (VASP) [36,37]. We used the projector-augmented wave (PAW) method with the Perdew-Burke-Ernzerhof (PBE) generalized gradient approximation (GGA) exchange-correlation functional [38,39]. A plane-wave energy cutoff of 520 eV was used. The Brillouin zone was sampled using a Monkhorst-Pack $k$-point grid with a density of $2\pi \times 0.025$ Å$^{-1}$. This mesh size is fine enough to sample the first Brillouin zone for achieving better $k$-point convergence [40]. The lattice vectors and the atomic positions of candidate structures selected from CGCNN predictions are fully optimized by the DFT calculations until forces on each atom are less than 0.01 eV/ Å. The exa-AMD framework enables the parallelization of these VASP calculations on multiple CPUs or GPUs, which significantly accelerates the DFT optimization. Depending on the elements of the system and size of the structures, we can complete DFT optimization of 5000 structures within 2 days in most cases, using 8 GPU nodes with 32 A100 GPUs.

**(v) Post-processing: formation energy, thermodynamic stability, magnetic calculations**
Once the total energies of all relaxed structures are obtained from DFT, the formation energy per atom for a Fe$_\alpha$Co$_\beta$Zr$_\gamma$ compound is calculated relative to the elemental phases using:

$$E_f = E(\text{Fe}_\alpha\text{Co}_\beta\text{Zr}_\gamma) - \alpha E(\text{Fe}) - \beta E(\text{Co}) - \gamma E(\text{Zr}),$$

where $E(X)$ is the total energy per atom of the reference elemental phase (e.g., bcc-Fe, hcp-Co).

While the formation energies $E_f$ are used for ML screening because these target values are better learned and predicted by CGCNN, the thermodynamic stability of a ternary compound is determined by the formation energy of the compound relative to the convex hull of the ternary system. For all the ternary compounds obtained after the DFT calculations, we also calculate their formation energy relative to the convex hull, $E_{hull}$, by DFT. This is done by comparing the formation energy of $Fe_\alpha Co_\beta Zr_\gamma$ with respect to the nearby three known stable phases. The chemical compositions of these phases are located at the vertices of the Gibbs triangle that encloses the composition of $Fe_\alpha Co_\beta Zr_\gamma$. We use this construction to assess the thermodynamic stability against decomposition into the stable phases. The $E_{hull}$ is the decomposition energy of a $Fe_\alpha Co_\beta Zr_\gamma$ ternary compound with respect to the nearby three known stable phases which can be ternary, binary, or elemental phases.

In this study, a stable structure at $T = 0$ K is defined as it has the formation energy on the convex hull ($E_{hull} = 0$ eV/atom), while a metastable structure is defined as it has a $E_{hull} \leqslant 0.1$ eV/atom above the convex hull. It has been shown by DFT calculations that about 50% of experimental synthesized structures in the Inorganic Crystal Structure Database [41] have $E_{hull}$ values above the convex hull and some structures can have $E_{hull}$ as high as 0.1 eV/atom [42]. Therefore, predicted structures with $E_{hull}$ below 0.1 eV/atom can be considered as experimentally synthesizable structures under non-equilibrium synthesis conditions.

Magnetic polarization, $J_s$, is a fundamental intrinsic property that quantifies the strength of a magnetic material. In our first-principles calculations, the total magnetic moment of the unit cell $M$ is determined by summing the individual atomic magnetic moments calculated by DFT at 0 K, representing a state of perfect magnetic ordering (saturation). The saturation magnetization, $M_s$, which is the magnetic moment per unit volume, is then calculated by dividing the total cell moment by the relaxed volume of the unit cell ($V_{cell}$): $M_s = M_{cell}/V_{cell}$.

The magnetic polarization $J_s$ is directly related to the saturation magnetization $M_s$ through the permeability of free space, $\mu_0$: $J_s = \mu_0 M_s$. Physically, $J_s$ represents the intrinsic magnetic flux density of the material, expressed in units of Tesla (T). It signifies the maximum possible density of aligned magnetic moments within the crystal structure. A high value of $J_s$ indicates that the material can generate a strong magnetic field and store a large amount of magnetic energy, which is a critical attribute for high-performance permanent magnets used in applications such as electric motors, generators, and data storage devices. Therefore, along with the magnetocrystalline anisotropy and Curie temperature, $J_s$ is one of the primary figures of merit we use to screen and identify promising new magnetic compounds.

For promising non-cubic structures with large magnetic moments, the magnetocrystalline anisotropy constant ($K_1$) are calculated by performing non-collinear DFT calculations including spin-orbit coupling. Initially, spin-polarized calculations for collinear magnetism are performed self-consistently. Subsequently, spin-orbit couplings are enabled, symmetry operations are turned off, and a non-self-consistent calculation is carried out [43]. The spin-quantization axis is set to the chosen direction. Magnetocrystalline anisotropy calculations are sensitive to fine crystal structure differences and parameter settings. We use a finer $k$-point mesh size of $2\pi \times 0.016 \text{Å}^{-1}$, and the convergence criterion of total energy is set to $10^{-8}$ eV, ensuring the calculated anisotropy energy is well converged. We calculate the total energy for magnetic moments oriented along the Cartesian (100), (010), and (001) directions, respectively. The direction associated with the lowest total energy is labeled as the magnetic "easy" direction, and the structure has uniaxial anisotropy. The direction with the second lowest total energy is labeled as the "intermediate" direction. The uniaxial magnetocrystalline anisotropy constant $K_1$ is defined as the total-energy difference, normalized by the volume of the unit cell [44]:

$$K_1 = (E_{intermediate} - E_{easy})/V.$$

If a whole plane has the lowest total energy (easy plane), with the perpendicular direction with higher energy (hard axis), then the structure has in-plane anisotropy, with $K_1$ defined as negative value:

$$K_1 = -(E_{hard} - E_{easy})/V.$$

This works for tetragonal and hexagonal structures. For orthorhombic structures, we also calculate directions like (110), (011) and (101), etc., ensuring our calculated $K_1$ is obtained properly. Note that it does not work for other systems where three axes are not orthogonal to each other, those structures are not considered here. A high uniaxial anisotropy is desirable for permanent magnet applications.

Within the framework of the Heisenberg model, the Curie temperature $T_c$ can be estimated using the mean-field approximation, which involves calculating the energy difference between the FM and AFM configurations. The mean field expression $T_c \sim \frac{2}{3} J_0$, where $J_0$ is the effective exchange interaction parameter evaluated from the difference $E_{AFM} - E_{FM}$ [45].

### III. RESULTS AND DISCUSSION

The 3102 structures selected from the CGCNN screening result in 531 distinct structures after the DFT optimization. Structures that cannot achieve self-consistent convergency in the DFT calculations are discarded. These discarded structures are most likely far from realistic Fe-Co-Zr ternary compounds. We then calculated $E_{hull}$, the formation energies of the 531 ternary compounds with respect to the ternary Fe-Co-Zr convex based on the total energies by DFT (see the above methods section for details).

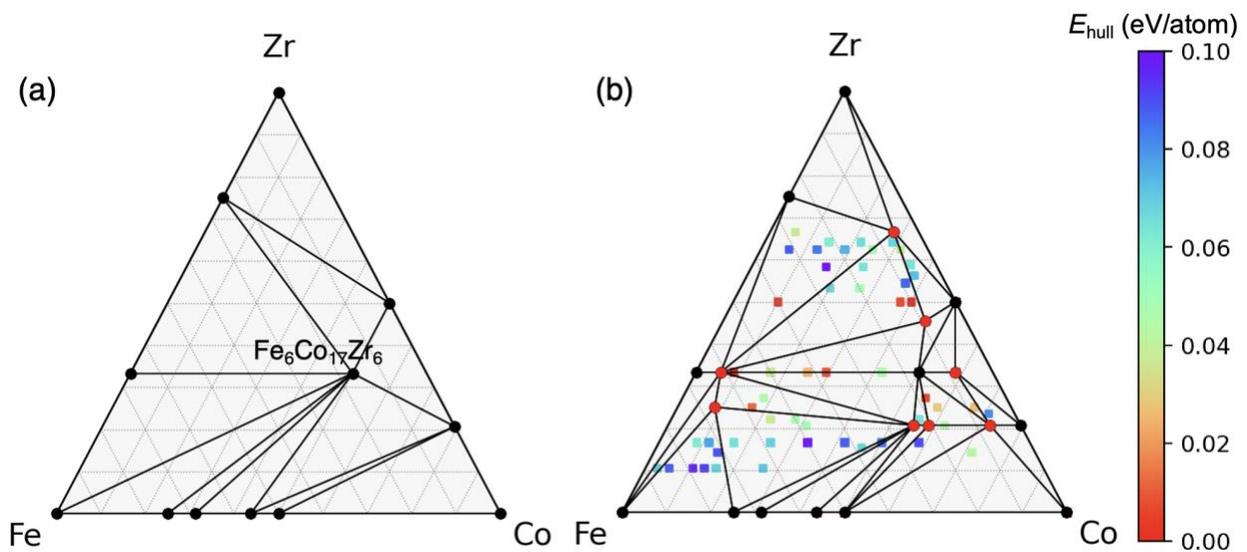

**Fig. 3.** (a) The existing convex hull of Fe-Co-Zr system with known stable elemental and binary phases, and ternary phase FeCoZr shown in black dots from Materials Project [32]. (b) The updated convex hull with the newly predicted Fe-Co-Zr compounds by DFT calculations based on the structures selected through exa-AMD framework. The red dots shown in (b) are newly predicted stable Fe-Co-Zr phases from our approach.

A major outcome of our ML-accelerated search is a significant update of the ternary Fe-Co-Zr phase diagram. Prior to this work, the thermodynamic landscape of this system was largely uncharted, with only one thermodynamically stable ternary compounds, $FeCo_3Zr_2$. This compound is included in the convex hull in the Materials Project database, as illustrated in **Figure 3(a)**. While several ternary Fe-Co-Zr compounds have been experimentally synthesized, including FeCoZr [9], $FeCo_3Zr_4$ (space group *Fd-3m*) [10], and $FeCoZr_4$ (space group *I4/mcm*) [11], our DFT calculations indicate that these phases are metastable and thus do not lie on the thermodynamic ground-state convex hull.

Our investigation dramatically enriches the ternary Fe-Co-Zr phase diagram with the discovery of 9 new, thermodynamically stable ternary compounds with formation energies on or below the previously known hull. It is important to note that three of these phases—$Fe_6Co_{17}Zr_6$, $Fe_7Co_{16}Zr_6$, and $Fe_2Co_{21}Zr_6$—were also predicted in the GNoME database [29]. This independent prediction validates the findings of both our exa-AMD framework and the GNoME database, confirming the reliability of our approach. The remaining six stable compounds are, to our knowledge, uniquely identified in this work.

These nine stable phases, represented by the red dots in **Figure 3(b)**, fundamentally redefine the ground-state energy landscape of the Fe-Co-Zr system. They establish new tie-lines and form new Gibbs triangles on the convex hull, indicating different decomposition pathways for other compositions. In addition to these stable ground states, our framework identified 81 unique metastable phases with low formation energies (within 0.1 eV/atom of the updated convex hull), which are also promising targets for non-equilibrium experimental synthesis. The compositions of these metastable structures are also shown in **Fig. 3 (b)**. More information of these metastable structures is given in the Supplementary Materials.

**Table 1.** The formula, space group, formula units per unit cell ($Z$), lattice parameters, and magnetic properties (magnetic polarization, $J_s$, magnetic anisotropy energy constant, $K_1$) of the 9 promising Fe-Co-Zr compounds discovered through our exa-AMD framework. Crystallographic data, such as lattice constants and atomic coordinates, of these compounds can be found in the Novomag Database [33].

| Formula | Space group | Z | Lattice parameters (Å) | | | $J_s$ (T) | $K_1$ (MJ/m³) |
| --- | --- | --- | --- | --- | --- | --- | --- |
| | | | a | b | c | | |
| Fe$_8$CoZr$_3$ (**Fig. 5(a)**) | $P6_3/mmc$ | 2 | 4.885 | 4.885 | 15.991 | 1.14 | -1.2 |
| Fe$_8$CoZr$_3$ (**Fig. 5(b)**) | $R$-$3m$ | 3 | 4.888 | 4.888 | 24.047 | 1.16 | -0.63 |
| Fe$_{11}$CoZr$_6$ (**Fig. 5(c)**) | $P6_3/mcm$ | 2 | 8.585 | 8.585 | 8.125 | 0.92 | 2.25 |
| FeCo$_5$Zr$_{12}$ (**Fig. 5(d)**) | $C222$ | 2 | 5.345 | 9.092 | 13.568 | 0 | 0 |
| FeCo$_7$Zr$_4$ (**Fig. 5(e)**) | $P$-$3m1$ | 1 | 4.914 | 4.914 | 7.926 | 0.15 | -1.11 |
| FeCo$_5$Zr$_5$ (**Fig. 5(f)**) | $Cmcm$ | 4 | 3.758 | 9.295 | 20.085 | 0.18 | 0.12 |
| Fe$_7$Co$_{16}$Zr$_6$ (**Fig. 5(g)**) | $C2/m$ | 2 | 14.048 | 8.147 | 8.164 | 1.05 | 0.05 |
| Fe$_6$Co$_{17}$Zr$_6$ (**Fig. 5(h)**) | $Immm$ | 2 | 8.158 | 8.167 | 11.59 | 1.02 | 0.60 |
| Fe$_2$Co$_{21}$Zr$_6$ (**Fig. 5(i)**) | $Immm$ | 2 | 8.145 | 8.16 | 11.48 | 0.93 | -0.26 |

In **Fig. 4**, we show the crystal structures of the 9 predicted new stable Fe-Co-Zr compounds. Some basic crystallographic information of these structures is also shown in **Table. 1**. The CIF files of these structures are given in the Supplementary Materials. The first 3 Fe-Co-Zr structures in **Fig. 4(a)-(c)** are Fe-rich with the ratio of 3d elements to Zr either 3:1 or 2:1, and only one Co atom in each formular unit. There are two stable polymorphs of Fe$_8$CoZr$_3$ (**Fig. 4(a) and (b)**). The first polymorph is in the hexagonal $P6_3/mmc$ space group, characterized by an elongated $c$-axis ($a$ = 4.885 Å, $c$ = 15.991 Å) and consisting of distinct Fe-Co and Zr atomic layers. The second polymorph (**Fig. 4(b)**) adopts a rhombohedral stacking sequence in the $R$-$3m$ space group, which results in a more complex layered arrangement and a significantly longer unit cell along the $c$-axis ($a$ = 4.888 Å, $c$ = 24.047 Å). The Fe$_{11}$CoZr$_6$ phase (**Fig. 6(c)**) adopts the $P6_3/mcm$ space group with lattice parameters of $a$ = 8.585 Å and $c$ = 8.125 Å, forming a complex three-dimensional framework with a pronounced layered character. These three compounds exhibit strong magnetization with $J_s$ value around or above 1 T. Unfortunately, the strong in-plane magnetic anisotropy makes them not suitable for permanent magnet applications.

The three structures in the middle row of **Fig. 4** are Fe-deficient. The FeCo$_5$Zr$_{12}$ (**Fig. 6(d)**) compound is a Zr-rich phase that crystallizes in the orthorhombic crystal system with space group $C222$. Its unit cell has lattice parameters of $a$ = 5.345 Å, $b$ = 9.092 Å, and $c$ = 13.568 Å. Given the high concentration of Zr, the structure consists of a complex framework of zirconium atoms, with the smaller iron and cobalt atoms occupying specific interstitial sites. Consistent with its high Zr content, this phase is calculated to be non-magnetic. The FeCo$_7$Zr$_4$ phase (**Fig. 6(e)**) adopts a trigonal structure with the space group *P3-m1* and lattice parameters of $a$ = 4.914 Å and $c$ = 7.926 Å. This compound is characterized by distinct atomic layers stacked along the *c*-axis, with Fe and Co atoms forming nets at specific heights and the Zr atoms intercalated between them. However, the magnetization of this structure is very small. The FeCo$_5$Zr$_5$ compound (**Fig. 6(f)**) crystallizes in the orthorhombic space group *Cmcm*. The unit cell is highly anisotropic, with lattice parameters of $a$ = 3.758 Å, $b$ = 9.295 Å, and a significantly elongated $c$ = 20.085 Å. A defining feature of this structure is its planar nature, with all atoms residing on crystallographic mirror planes, forming corrugated layers that are stacked along the *c*-axis. The magnetization of this structure is also small, unsuitable for permanent magnets applications.

Finally, the remaining three stable compounds shown in the bottom row of **Fig. 4** are closely related, all sharing a common stoichiometry of approximately Zr$_6$X$_{23}$ (X = Fe or Co). Two of these phases, Fe$_6$Co$_{17}$Zr$_6$ (**Fig. 4(h)**) and Fe$_2$Co$_{21}$Zr$_6$ (**Fig. 4(i)**), are isostructural, both in the body-centered orthorhombic space group *Immm* with very similar lattice parameters. They represent variations where the Fe and Co atoms are distributed differently across the same underlying atomic framework. The third phase, Fe$_7$Co$_{16}$Zr$_6$ (**Fig. 4(g)**), can be considered as a monoclinic distortion of the same structural motif in **(h)** and **(i)**. It adopts the space group *C2/m* with a characteristic monoclinic angle of $\beta$ = 125.4°. The discovery of these three distinct but related stable phases suggest the robustness of the Zr$_6$X$_{23}$ structural archetype across a range of Fe/Co ratios. These three structures exhibit similar magnetic polarization $J_s$ but their magnetic anisotropy is very different. Partially switching Fe and Co, or even adopting disordered alloy structures with partial occupation, would further lower the free energy, but the disorder in the Fe-Co would also strongly affect magnetic anisotropy. This will be an interesting topic for future investigations.

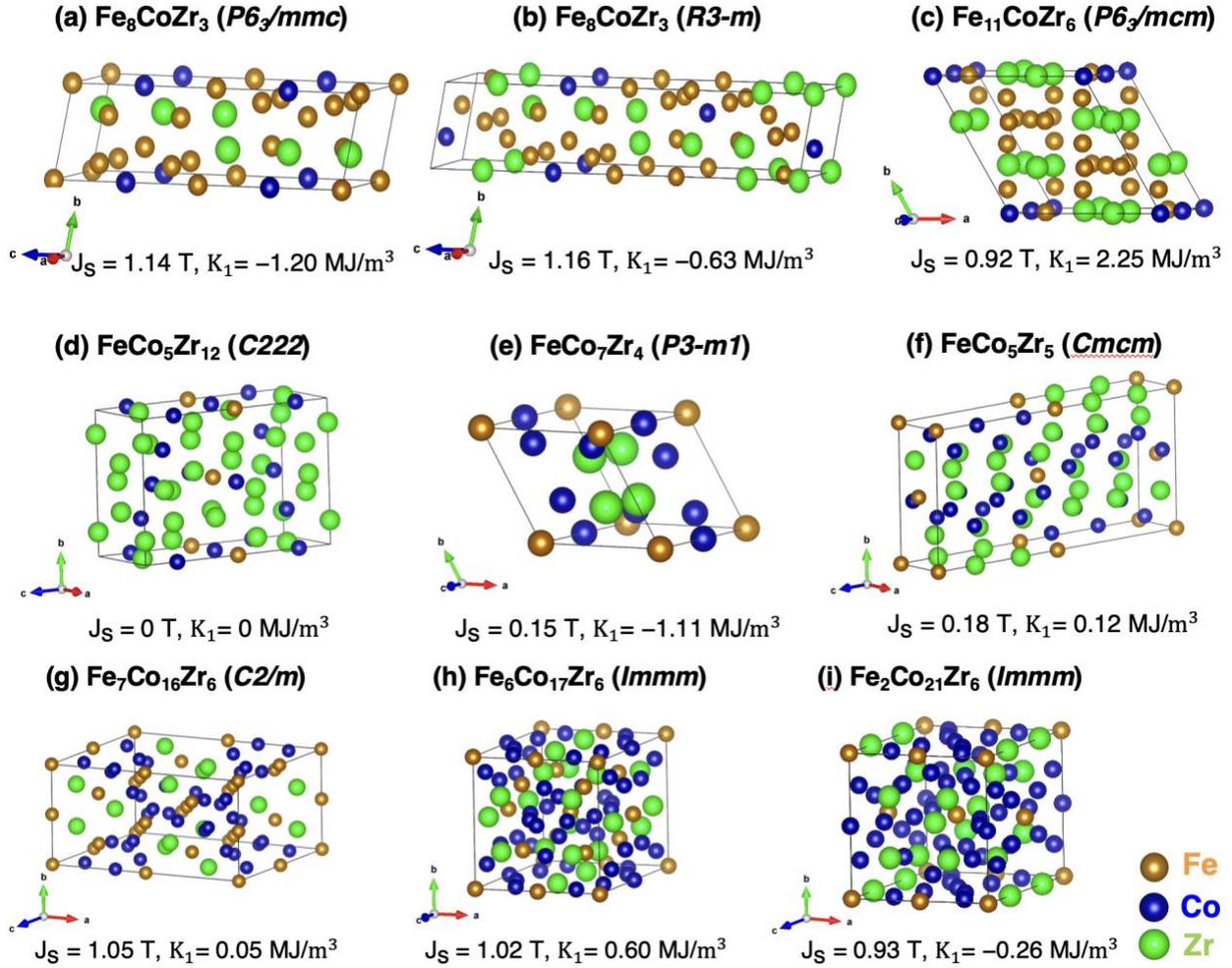

**Fig. 5.** The crystal structures of the 9 Fe-Co-Zr ternary compounds that are predicted to be stable.

We note that among the 9 stable structures obtained from our ML-assisted search, only the Fe6Co17Zr6 phase (**Fig. 4 (h)**) has some potential for RE-free permanent magnetic applications since it has decent magnetization ($J_s$ = 1.02 T) and uniaxial magnetic anisotropy ($K_1$ =0.6 MJ/m$^3$, higher than that of *hcp* Co which is 0.41 MJ/m$^3$). To gain further insight on contributions from different atoms to the magnetic anisotropy in this compound, we performed a site-resolved analysis of its magnetocrystalline anisotropy energy (MAE).

The total MAE can be considered as a summation of contributions from each atomic site, where the site-resolved MAE ($K_{SOC}$) for each atom reveals its impact on the overall anisotropy. We decompose the total MAE into contributions from different atomic sites, following the spatial decomposition scheme of Antropov et al [46]. Here, the MAE is expressed as half of the sum of the atomic spin-orbit coupling anisotropies over all atomic sites in the unit cell:

$K = \frac{1}{2}\sum_i K_{SOC}(i)$

In this scheme, the site-resolved MAE (i.e., $K_{SOC}(i)$) is defined as the difference in the spin-orbit coupling energy between two magnetization orientations at site $i$. By examining this site-resolved MAE, we found that both Fe and Co atoms contribute significantly to the total MAE, with contributions much higher than that of *hcp* Co.

Our analysis of $Fe_6Co_{17}Zr_6$, detailed in **Table 2**, revealed that the two Fe atoms located at the *2a* Wyckoff sites and two Co atoms at *2c* Wyckoff sites provide a significant but negative contribution to the uniaxial anisotropy, with a $K_{SOC}$ of -0.788 meV and -0.736 meV per atom, respectively. This finding provided a clear, data-driven hypothesis: selectively replacing these specific Fe/Co atoms, which actively oppose the desired magnetic hardness, could significantly improve the compound's overall $K_1$.

**Table 2.** Site-resolved spin-orbit coupling anisotropy energies for $Fe_6Co_{17}Zr_6$.

| Atom | Site | # Multiplicity | $K_{SOC}$ (meV) | Moment ($\mu B$) |
|---|---|---|---|---|
| Zr | *n* | 8 | 0.064 | -0.293 |
| Zr | *j* | 4 | -0.078 | -0.294 |
| Fe | *k* | 8 | 0.420 | 2.345 |
| Fe | *a* | 2 | **-0.788** | 2.349 |
| Fe | *b* | 2 | 0.020 | 2.687 |
| Co | *l1* | 8 | 0.008 | 1.177 |
| Co | *l2* | 8 | -0.065 | 1.313 |
| Co | *m1* | 8 | 0.286 | 1.184 |
| Co | *m2* | 8 | 0.214 | 1.320 |
| Co | *c* | 2 | **-0.736** | 1.433 |

Acting on this hypothesis, we performed a targeted computational experiment by substituting the two Fe atoms at the *2a* sites with Co atoms. This local substitution successfully transformed the parent structure, which has orthorhombic *Immm* symmetry, into a new phase, $Fe_5Co_{18}Zr_6$, with an increased tetragonal *I4/mmm* symmetry, as illustrated in **Figs. 5 (a)** and **(b)**. Our DFT calculations show this new tetragonal structure is a highly accessible near-stable phase, with a formation energy of only 1 meV/atom above the convex hull. It is noteworthy that this *I4/mmm* phase is 43 meV/atom lower in energy than a $Fe_5Co_{18}Zr_6$ structure with *C2/m* symmetry previously predicted by the GNoME database [29], highlighting our capability to identify novel, lower-energy structural configurations.

**Table 3.** Site-resolved spin-orbit coupling anisotropy energies for $Fe_5Co_{18}Zr_6$ with Fe *a* sites substituted by Co atoms.

| Atom | Site | # Multiplicity | $K_{SOC}$ (meV) | Moment ($\mu B$) |
|---|---|---|---|---|
| Zr | *h* | 8 | 0.129 | -0.284 |
| Zr | *e* | 4 | -0.121 | -0.282 |
| Fe | *f* | 8 | 0.539 | 2.342 |
| Fe | *b* | 2 | 0.035 | 2.669 |
| Co | *n1* | 8 | 0.162 | 1.188 |

| Co | n2 | 8 | 0.029  | 1.298 |
| Co | n3 | 8 | 0.246  | 1.187 |
| Co | n4 | 8 | 0.207  | 1.299 |
| Co | c1 | 2 | 0.401  | 1.436 |
| Co | c2 | 2 | -0.546 | 1.436 |

Intriguingly, this precise atomic substitution resulted in a dramatic enhancement of magnetic anisotropy. The calculated $K_1$ for $Fe_5Co_{18}Zr_6$ is 1.1 MJ/m³, an 83% improvement over the 0.60 MJ/m³ of the original $Fe_6Co_{17}Zr_6$ compound. The site-resolved analysis for $Fe_5Co_{18}Zr_6$ (**Table 3**) shows that replacing the Fe atoms at the $2a$ sites with Co atoms (now at $2c1$ sites) mitigates the strong negative contribution to the anisotropy. This targeted modification demonstrates that fine-tuning the local atomic environment is a powerful strategy for optimizing the magnetic properties of ternary intermetallics.

We further extend this site-substitution strategy by replacing the Fe atoms at the $2a$ sites and the Co atoms at the $2c$ sites with Manganese (Mn), leading to the creation of a new, thermodynamically stable quaternary phase, $Fe_5Co_{16}Zr_6Mn_4$ (**Fig. 5(c)**). While this substitution achieves thermodynamic stability, it is detrimental to the magnetic anisotropy, which decreases to 0.35 MJ/m³. This result serves as an important counterpoint, illustrating the delicate balance between achieving stability and enhancing magnetic properties, and underscores the value of a guided, analysis-driven approach to material design. This targeted optimization, which builds upon the success of the initial broad search, is a powerful and efficient method for rationally engineering next-generation magnetic materials.

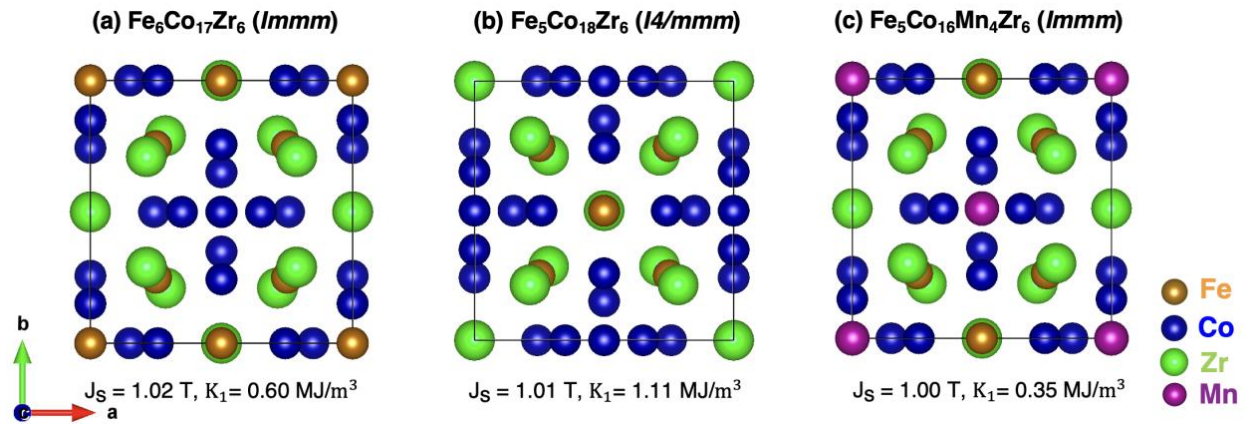

**Fig. 6.** Crystal structures of the original predicted (a) $Fe_6Co_{17}Zr_6$, the derived stable phases (b) $Fe_5Co_{18}Zr_6$, and (c) $Fe_5Co_{16}Zr_6Mn_4$ through local substitutions.

We also calculate the Curie temperatures for these three compounds by mean field approximation within the Heisenberg model as described above. **Table 4** summarizes the calculated saturation magnetization ($J_s$), magnetic anisotropy constant ($K_1$), and estimated Curie temperature ($T_c$) for these three compounds. The estimated Curie temperatures for these candidates are well above room temperature, which is essential for practical applications. These results provide promising candidates for experimental synthesis and characterization.

**Table 4.** Formula, space group, and magnetic properties (magnetic polarization, $J_s$, magnetic anisotropy energy constant, $K_1$, magnetic easy axis, and Curie temperature, $T_c$) of the 3 promising Fe-Co-Zr compounds through our approach. For comparison, experimental data for *bcc* Fe and *hcp* Co are given in italic.

| Formula | Space group | $J_s$(T) | $K_1$ (MJ/m$^3$) | easy axis | $T_c$(K) |
| --- | --- | --- | --- | --- | --- |
| Fe | $Im\bar{3}m$ | 2.15 | | | *1043* |
| Co | $P6_3/mmc$ | 1.81 | 0.41 | c | *1388* |
| Fe$_6$Co$_{17}$Zr$_6$ [Fig. 5(a)] | $Immm$ | 1.02 | 0.60 | c | 1547 |
| Fe$_5$Co$_{18}$Zr$_6$ [Fig. 5(b)] | $I4/mmm$ | 1.01 | 1.11 | c | 1648 |
| Fe$_5$Co$_{16}$Zr$_6$Mn$_4$ [Fig. 5(c)] | $Immm$ | 1.00 | 0.35 | c | 1311 |

## IV. SUMMARY

In summary, we have performed a comprehensive and systematic search for novel magnetic compounds in ternary Fe-Co-Zr system using a scalable and efficient ML-guided discovery framework. This integrated approach, which combines the exa-AMD workflow for high-throughput screening and first-principles calculations, has dramatically accelerated the exploration of the vast chemical and structural space.

Our search successfully identified **9** new thermodynamically stable and **81** low-energy metastable Fe-Co-Zr compounds, significantly advancing the knowledge of this ternary system. Furthermore, we demonstrated a powerful strategy for property optimization through targeted local atomic substitution, enhancing the magnetic anisotropy of a stable phase Fe$_6$Co$_{17}$Zr$_6$ to create a promising candidate, Fe$_5$Co$_{18}$Zr$_6$. The promising magnetic properties calculated for several of these newly discovered compounds make them compelling candidates for RE-free permanent magnets. Moreover, the local substitution approach led to another stable quaternary phase, Fe$_5$Co$_{16}$Zr$_6$Mn$_4$. This highlights the capability of our approach to navigate complex chemical spaces towards predictions of new stable phases, as well as engineering of tunable magnetic properties such as magnetic anisotropy.

The primary advantage of our exa-AMD framework over traditional crystal structure search methods is its exceptional computational efficiency. The CGCNN screening process evaluated the formation energies of 884,550 hypothetical compounds in just 10 minutes on a standard GPU, efficiently narrowing the vast search space to 3,102 viable structures (~0.35% of the total) spanning over 1,723 compositions for further DFT analysis. This efficiency extends to the most computationally intensive stage; the framework's ability to parallelize DFT calculations across multiple GPUs enabled the full first-principles evaluation of all 3,102 candidates to be completed in less than 12 hours on a modest cluster of 16 GPU nodes (64 A100 GPUs total). This level of throughput allows for a much wider and more rapid exploration of the compositional space than is feasible with traditional search algorithms. Nevertheless, it is important to note that our substitutional approach does not guarantee exhaustive coverage, as it is dependent on the structural templates present in the initial databases and may overlook novel motifs not represented in the training data.

This work provides concrete, computationally-vetted targets that can help experimentalists bypass much of the time-consuming trial-and-error optimization process. Future work should focus on the experimental synthesis and characterization of the most promising phases identified here, as well as computational investigations into disordered structures with partial atomic occupancy. The success of our approach validates this ML-guided framework as a robust paradigm for materials design that can be readily extended to other complex ternary and quaternary systems. It represents a powerful tool for rapidly identifying novel materials, marking a significant step forward in the digital era of materials discovery and design.

**Acknowledgments**
Work at Ames National Laboratory and Los Alamos National Laboratory was supported by the U.S. Department of Energy (DOE), Office of Science, Basic Energy Sciences, Materials Science and Engineering Division through the Computational Material Science Center program. Ames National Laboratory is operated for the U.S. DOE by Iowa State University under contract # DE-AC02-07CH11358. Los Alamos National Laboratory is operated by Triad National Security, LLC, for the National Nuclear Security Administration of U.S. Department of Energy under Contract No. 89233218CNA000001. This research used resources of the National Energy Research Scientific Computing Center (NERSC), a DOE Office of Science User Facility supported under Contract No. DE-AC02-05CH11231. J.R.C. acknowledges support from the 2025 Hill Prizes on Physical Sciences.